\documentclass[prl,aps,epsf]{revtex4}
\usepackage{graphicx}
\usepackage{amsmath}
\usepackage{amssymb}

\begin{document}
\draft                         
\pagestyle{myheadings} \markright{{\sf Lansac et al.}: {\sl Discrete 
elastic model for two-dimensional melting}}
%

\title{Discrete Elastic Model for Two-dimensional Melting}

\author{Yves Lansac, Matthew A. Glaser and Noel A. Clark}
\address{Condensed Matter Laboratory, Department of Physics, and \\
         Ferroelectric Liquid Crystal Materials Research Center, \\
         University of Colorado, Boulder, CO 80309, USA}

\date{}                              


\begin{abstract}

Our previous molecular dynamic simulation studies of simple two-dimensional
(2D) systems \cite{matt_big} suggested that both geometrical defects 
(localized, 
large-amplitude deviations from hexagonal ordering) and topological defects
(dislocations and disclinations) play a role in 2D melting. To capture
the main features of the 2D melting transition and investigate the respective
roles of these two classes of defects, we study a discrete elastic model
consisting of an array of nodes connected by springs, for which the relative
number of geometrical defects (modeled as broken springs)
and topological defects (nodes having a coordination number
different from 6) may be precisely controlled. We preform Monte Carlo 
simulations of this model in the isobaric-isothermal ensemble, and present
the phase diagram as well as various thermodynamic, statistical and 
structural quantities as a function of the relative populations of geometrical 
and topological defects. The model exhibits a rich phase behavior including
hexagonal and square crystals, expanded crystal, dodecagonal quasicrystal, 
and liquid structure. We found that the solid--liquid transition temperature is
lowered by a factor $3$ to $4$, with respect to the case when only 
topological defects are allowed, when both topological and geometrical
defects are permitted, supporting our hypothesis concerning the central
role played by the geometrical defects in the melting.
The microscopic structure of the dense liquid has been investigated and
the results are compared to those from simulations of 2D particle systems. 

\end{abstract}

\maketitle
%


\section{Introduction}
The process of melting is one of the most common  yet
fascinating phenomena encountered in every day life.
The abrupt changes in the properties of a substance when
it is heating above its melting point, i.e. the dramatic 
change from solidity through fluidity, achieved with very
small change in density, remain mysterious and intriguing.
If modern statistical mechanics theory, through the integral equation 
\cite{hansen} approach, allow to compute, although with approximation, the
pair correlation function and the thermodynamics properties
of dense liquid, it does not provide an intuitive microscopic
picture of a dense fluid nor give an understanding of the entropy
of a dense fluid or the entropy difference between the solid and
liquid states.

In the 60's, a crystallographer, J.D. Bernal, tried to relate the
structure of a liquid, not too far away from its freezing point, to 
excluded volume problem \cite{bern1,bern2,bern3,bern4}. 
In Bernal's view, a liquid is essentialy
homogeneous, coherent and irregular in opposite with a crystal which
may be seen as a pile of molecules regularly arranged.
His work consisted to determine the local organization of molecules
in various mechanistic model governed by geometrical packing constraints
and to identify the kind of polygons defined by the bonds binding the
center of mass of neigbors particles, forming holes 
responsible of the less packing organization in a dense
liquid in contrast to the close packing of solids. His approach was 
primarily descriptive, and was almost completely abandonned in favor 
of the modern integral equation approach.
However, the present description shear a lot of idea with these earlier
works because we think that this approach is an important way to 
develop a general description of melting theory and to provide a very
intuitive picture of the structure and properties of a dense liquid. 

In addition to the general understanding of the liquid structure, the
two-dimensional (2D) case is even more puzzling due to the fact that 
no true long-range
positional order is present \cite{mermin1,mermin2,mermin3}. 
Different theories predict that this
difference in the nature of the 2D and 3D solid lead to a completely
different melting mechanism.
In particular, the theory developed by Kosterlitz and Thouless 
\cite{kost1,kost2} and extended 
by Halperin, Nelson \cite{halp1,halp2} and
Young \cite{young} (KTHNY) has received a lot of attention.
According to their prediction, the solid-liquid transition in two dimensions
occurs by the intermediate of two second order phase transitions, corresponding
to the unbinding of dislocations and disclinations. In addition, a new phase
called the hexatic phase is predicted to appear between the solid and the 
liquid. This phase is characterized by short-range positional order and by
quasi long-range bond orientational order.
A lot of work, both numerical and experimental have been done in order to 
probe this theory. However no strong evidence for the KTHNY theory has 
emerged from the theoretical studies, and the results from most of the 
simulations
indicate that another melting mechanism preempts the KTHNY mechanism.

Our previous molecular dynamics (MD) simulations \cite{matt_big} performed 
in the microcanonical ensemble and based on a system of disks interacting via
a soft repulsive potential have focused on the mechanism
of melting and were devoted to a better intuitive understanding of the
microscopic structure of a dense liquid. The results have shown that the
melting occurs via a first-order phase transition and that an
important aspect of the transition is the appearence of geometrical
(non topological) disorder in the liquid phase. This disorder manifests itself 
in the presence of a significant degree of square lattice coordination
(holes, see Figure~\ref{fig:2d_cartoon}) in the liquid phase with particles 
adopting local arrangments characteristic of plane tilings composed of squares 
and equilateral triangles (ST tilings)\cite{collins,kawa,yi}. 
Dense random packings (DRPs) \cite{ziman,finney} exhibit 
the same kind of ST tiling like structures and are quite supportive of the 
fact that geometrical constraints play an important role in melting and 
the liquid structure.
Topological defects are responsible for the loss of long-range positional
order characteristic of a liquid, but the geometrical excitations, not 
captured by the Voronoi \cite{voronoi} or Delaunay 
(the dual of the Voronoi representation)
representations, formed by almost local square organization
of the particles may contribute significantly to the thermodynamics
(i.e the change in density) occuring at the melting transition.  
Both topological {\sl and} geometrical defects are the 
fundamental excitations which control the transition and formed the basis
the basis of our mechanistic picture of a dense liquid:
the two-dimensional melting is a condensation
of localised, thermally generated geometrical and topological defects. 
Condensation is a phase
transition resulting from attractive interactions between the excitation or
particles of interest, i.e. in our case between the defects.
The scalar order parameter of the transition (i.e. the number density
of geometrical defects) is unrelated to the symmetry-derived order parameters
and then the global symmetry change could be very well seen as a side effect
of the defect condensation mechanism which drives the 2D melting transition.
This picture is similar to the one obtained in a three-dimensional liquid by 
the mechanistic models of Bernal. This is not too surprising, because in
the framework of a melting transition driven mainly by packing constraints,
global symmetry is irrelevant in determining the chracteristics of the melting
transition, then there is no compelling reason that 3D melting should
be qualitatively different from 2D melting.

Based on these observations, we present here a discrete elastic model 
constituted by a network of nodes connected together by elastic springs 
which capture the essential features of the structure of a liquid
not too far from the freezing point. 
This model can be seen as an analog of the mechanistic models from Bernal, 
modern computer and Monte-Carlo techniques allowing to have a much more
powerful tool to probe the statistical structure of liquids.


In the first part of the article, we describe the discrete elastic model and 
the numerical methodology applied in order to mimic a dense liquid.

In the second part we present the phase diagrams obtained with the model
as a function of the relative populations of the two different kind of 
excitations. Then the structure of the dense liquid is discussed
in detail and various thermodynamical and structural 
quantities are computed and compared with our earlier simulations, 
hereafter referred as WCA liquid.

In conclusion we discuss
the pertinence and limitations of our model as well as the possible extensions
and applications, in particular to freely suspended liquid crystal thin films.

\section{Discrete Elastic Model}

From the WCA liquid study, the 2D melting can be seen as
a defect condensation transition involving both geometrical and 
topological defects. 

Topological defects (disclinations) are characterized by a number of
nearest-neighbors different from 6 (i.e, different from what is expected
for a perfect triangular lattice). A node (playing in our model the role
of the center of mass of a disklike particle) having 5
nearest-neighbors is identified as a -1 disclination due to the fact that it
corresponds to the removal of a 60$^o$ wedge of material at a microscopic 
level, while a node having 7 nearest-neighbors is identified as a +1 
disclination, corresponding to the addition of a 60$^o$ wedge of material 
at a microscopic level. An isolated disclination corresponds to the 60$^o$ 
rotation of a vector
oriented along one of the lattice direction upon parallel transport around
a closed circuit enclosing the disclination and is responsible for the loss
of long-range bond-orientational order in the system.
An isolated dislocation is formed by a bound pair of disclination of opposite
strength and corresponds to a non-zero Burger vector (the amount by which a
Burger's circuit fails to close around a dislocation) and is responsible for
the loose of quasi-long-range positional order in the system.
In the liquid phase, a Burger's circuit is a ill-defined quantity since no
lattice is present.

Geometrical defects are non-topological defects and large amplitude 
localized geometrical distorsions in which the particles have adopted 
a nearly square organization but are not associated with a disclination
or dislocation (see Figure~\ref{fig:2d_cartoon}).
There is a clear tendency for particles in the dense 2D WCA liquid to form
local arrangments characteristic of ST tilings. 
The WCA liquid can be usefully describe as polygon packings or imperfect
tilings (i.e. formed with deformable tiles), and the effect of packing
constraints on the local geometry of these dense systems can be embodied in
particular tiling rules that condition the local arrangment of the polygons. 
The volume increase upon melting is directely related to the creation of 
polygons having more than 3 sides. 
 
The discrete elastic model is built to mimic the polygon tiling 
representation obtained from the Delaunay representation in 
the WCA liquid and to reproduce the two kinds of excitations 
responsible for the melting transition and the dense liquid structure. 
It is constituted by a network of nodes 
(playing the role of the center of mass of the disk particles of the WCA 
liquid) connected by springs. 
In addition to the bond stretching springs, bond angle bending
springs have been considered in order to avoid unphysical 
folding of the network at high pressure (Figure~\ref{fig:2d_model}).
A solid-like local structure is characterized by a triangular equilateral
network of springs connecting nodes, or equivalently by a plane tiling of
nearly perfect equilateral triangles. 
One of the main interests of the model
over its real-particle counterpart is that it is possible to probe almost
independently the effect of topological and geometrical excitations on the
melting transition and the resulting dense liquid structure.

Within the framework of the discrete elastic model, the creation of 
topological defects is reproduced by a local change in the connectivity of 
the network achieved by the flipping of a given spring bond between 
two nodes. The network topology is locally modified, resulting in the creation 
of a bound-pair of dislocations (Figure~\ref{fig:2d_moves}a).
Such a neutral quadrupole arrangment does not disrupt the long-range positional 
order of the system, but is at the origin of the subsequent creation
of topological defects by unbinding mechanism. 
The creation of isolated dislocations then isolated disclinations from such a 
quadrupole is schematically depicted on Figure~\ref{fig:2d_moves}a. 

The creation of a geometrical defect is achieved by a bond breaking
(Figure~\ref{fig:2d_moves}b). This procedure allows the two previously
connected nodes to remain in the 
proximity of their former nearest-neighbors without being strongly bound
to them, behaving as nearly as second-neighbors. This is a weaker 
effect than the change of neighborood induced by a bond flipping.   
Practicaly, the bond breaking procedure is accompanied
by an update of the equilibrium bond angles to reflect the induced change
in geometry from two equilateral tiles (bond angle springs set to 60$^o$)
to one nearly-square tile (bond angle springs set to 90$^o$)
(Figure~\ref{fig:2d_moves}b).

These two procedures (bond flipping and bond breaking) are suitable to 
mimic a system in which the populations of the two kinds of defects
can be controlled and their resulting effects on the melting transition
analysed. If only bond flipping is allowed, we will talk about an only
topological model, if only bond breaking is allowed, the model will be 
qualified of only geometrical model and if both bond flipping and bond
breaking are permitted, we will talk of a topological and geometrical
model. 
It is worth noticing, however, that local, square-like fluctuation of the 
triangular lattice are also present in the only topological model, and could be 
exhibited by using a bond dilution procedure ({\sl i.e} removal of all the 
bonds significantly longer than the average bond length present in the 
network). However, these fluctuations are, due to the constraints imposed on 
the model, much less favorable energetically than in the model where bond 
breaking is allowed.  This is in this sense that a distinction has been made 
between the only topological model and the topological and geometrical model.

Our model can handle the creation of geometrical defects up to six sides 
polygons (hexagonal tiles)
and the creation of topological defects of different strength ({\sl i.e.} with 
different coordination numbers) but in this article we focus on the 
minimal model in which only triangular and square tiles 
(3 and 4-sides polygons) and topological defects with 5 and 7 coordination 
numbers are considered. This is in agreement with the extensive studies 
performed on the WCA liquid and the random close packing system in which 
it was demonstrated that these excitations are the most relevant ones.

The Hamiltonian for this system is:

\begin{equation}
H = \frac{K_r}{2} \sum_{i,j} {({\bf r}_{i,j} - {\bf a})}^{2} +
    \frac{K_{\theta}}{2} \sum_{i,j,k} {({\theta}_{i,j,k} - {\theta}_{p})
}^{2} + P A 
\end{equation}

where the first sum is on all non-broken bonds and the second on all the 
bond-angle between non broken bonds, $|{\bf a}| \equiv a$ is the equilibrium
length between the two connected nodes $i$ and $j$ and ${\theta}_{p}$
is the equilibrium bond angle between nodes $i$, $j$ and $k$, for the polygon 
$p$ considered (${\theta}_{p}$ = 60$^{o}$, for $p$ = 3 {\sl i.e.} for an
equilateral triangle polygon and ${\theta}_{p}$ = 90$^{o}$, for $p$ = 4 
{\sl i.e.} for a square polygon).
In the case when both topological and geometrical defects are allowed
no energetic cost is associated with the flipping of a broken bond.
That is related to the fact pointed above, that a quadrupole of defects 
is nearly topologically equivalent to no defects. 
$K_{r}$ is the stretch elastic constant, $K_{\theta}$ the bend elastic 
constant, $A$ the area of the system and $P$ the external pressure.

The properties of the elastic model depend only on three dimensionless
parameters, a reduced temperature $t$, a reduced elastic constant ratio $K$
and a reduced pressure $p$ defined as:

\begin{eqnarray}
t & = & \frac{k_{B}T}{K_{r} a^{2}} \\ \nonumber \\
K & = & \frac{K_{\theta}}{K_{r} a^{2}} \\ \nonumber \\
p & = & \frac{P}{K_{r} a^{2}} 
\end{eqnarray}

where $k_{B}$ is the Boltzman constant and $T$ the temperature of the
system.

The statistical mechanical properties of the discrete elastic model are
investigated using Metropolis Monte Carlo (MC)\cite{metro} simulations
carried out in the isobaric-isothermic ensemble.  
Translational node displacements are also considered in order to relax the 
local strain induced in the system by the creation of defects resulting 
in a network connectivity change.  In addition, the simulation box area 
is allow to fluctuate in order to insure a constant pressure in the system.

\section{Phase diagrams}

In order to check the ability of our model to reproduce the solid-liquid
transition and in order to probe the relative importance of the two kind
of excitations in our model, we compute the $(p,t)$ phase diagrams
corresponding to the cases when only geometrical, only topological,
both geometrical and topological defects are allowed. A system
with $N=100$ nodes and then $3 N$ bonds is used. Additional simulations
have been carried out for selected state points with a larger system
with $N = 1024$ nodes in order to assess finite size effects and their 
influence on the location of the phase boundaries. 
A value $K=0.1$ has been chosen for all the considered models.
The effect of an increase of the elastic network stiffness will result in an 
increase in the liquid-solid transition temperature and its effects on the 
location of other phase boundaries will be study in another work.

At each state point ($p,t$), a Monte-Carlo sampling is used corresponding to
$200 000$ sweeps for equilibration and 10$^6$ sweeps for production. The 
rough location of the phase boundaries is obtained upon heating the system by 
successive steps $\Delta t$. Smaller steps in temparature are subsequently
used in order to refine the phase boundary location.   
Every sweep consist in the attempting move of the N nodes,
attempting move of the $3 N$ bonds (i.e. only fliping in the case when only
topological defects are considered, only breaking when only geometrical 
defects are considered and either breaking or fliping when both kind of
excitations are presents) and one attempting  move of the simulation box
(either only the area, or only the shape or a combination of the two 
precedent kind) in order to keep a pressure fluctuating around the specified 
value.

For each pressure, the location of the phase transition as a function of 
the reduced temparature, and the structure of the phases are determined 
by computing the equation of state, the energy, the specific heat and 
the magnitude of the bond-orientational order parameters.
Various bond-orientational order can be characterized  
by considering, in a general way, a bond angle density for a particle 
(node) $i$ defined as;

\begin{equation}
{{\rho}_b}_i(\theta) = \frac{2 \pi}{n_i} \sum_{j = 1}^{n_i}
\delta(\theta - {\theta}_{ij})
\end{equation}

where the sum runs over the $n_i$ nearest-neighbors of particle $i$ and
$\theta_{ij}$ is the angle of the bond between particle $i$ and its $jth$
neighbor with respect to a reference direction.
Because ${{\rho}_b}_i(\theta)$ has a periodicity of $2 \pi$, we can write
a Fourier decomposition:

\begin{equation}
{{\rho}_b}_i(\theta) = \sum_{m = 0}^{\infty} {{\psi}_m}_i \exp(-i m \theta)
\end{equation}

where the complex Fourier coefficients ${\psi}_m$ are given by:
\begin{equation}
{\psi}_m \equiv {{\psi}_m}_i = \frac{1}{n_i} \sum_{j = 1}^{n_i}
\exp(i m {\theta}_{ij})
\end{equation}

plays the role of a m-fold local bond orientational order parameter.
In addition we can define a global m-fold order parameter:

\begin{equation}
{\Psi}_m = \frac{1}{N} \sum_{i = 1}^{N} {\psi}_m
\end{equation}

which gives a signature of the global m--fold order present in the system.
We focus here on ${\Psi}_{4}$, ${\Psi}_{6}$ and ${\Psi}_{12}$ 
which are respectively sensitive to squarelike solid phase, triangular
solid phase and dodecagonal quasicrystalline phase.
A liquid phase will be characterized by negligible magnitudes of these
three order parameters. 

The resulting phase diagrams corresponding to the only geometrical model,
the only toplogical model and the topological and geometrical model are
presented respectively on Figure~\ref{fig:2d_diag}a, b and c.

When only geometrical defects are allowed (Figure~\ref{fig:2d_diag}a)
no change is symmetry is expected. 
The system exhibits an hexagonal crystal phase (X$_h$), an expanded hexagonal 
crystal (EX) i.e. an hexagonal crystal with a lower density to the presence of 
a significant number of geometrical defects (square-like fluctuations) and a 
quasicrystalline (QX) phase (dodecagonal, formed by perfect ST tiling 
\cite{chen,leung,yang,kuo}).
A triple point (X$_h$, QX, EX) is located around $p \simeq 0.03$ and $t \simeq
0.0068$. 
The crystal-expanded crystal transition is similar to a liquid-gas phase 
transition and therefore is first order and end by a critical point located 
around p = 0.055.
The location of this phase boundary has been characterized by the apparition
of a sharp peak in the variation of the specific heat with the temperature. 
A representative configuration of the expanded crystal is shown on 
Figure~\ref{fig:2d_snap}.

The location of the critical point has been achieved by performing additional 
simulations with the N = 1024 system at a pressure $p = 0.05$ and $p = 0.07$.
At this latter pressure no discontinuity in the order parameters or peak in
the specific heat is noticeable.
At pressures lower than $p = 0.03$ a quasicrystalline phase is stabilized.
The relative stability of the  hexagonal crystal phase (i.e a perfect 
tiling of the plane with equilateral triangular tiles) and the dodecagonal 
quasicrystalline phase (i.e. a perfect tiling of the plane with equilateral 
triangular and square tiles) arises as a competition between the 
entropy of configuration (the different arrangement of the tiles) - and to 
a lesser extend the entropy of vibration of the nodes - and the enthalpy
(the volume change). At low enough pressure, the gain in entropy induced
by a quasicrystalline phase overcomes the increase in volume for a given
range of temperature ($0.005 < t < 0.0071$, at $p = 0$), stabilizing the QX 
phase. 
As expected the range of stability of the QX phase with respect to the crystal 
phase decreases when the pressure increases.
As shown for $p = 0.02$, finite-size effects seem to increase the range of 
stability (vs. the crystal phase). A representative configuration (p = 0.02
and t = 0.0064) of the dodecagonal quasicrystalline phase is shown on 
Figure~\ref{fig:2d_snap2}. The number of square polygons divided by the
number of triangle polygons is $\sim 0.44$ and is constant within the full
temperature range ($0.006 < t < 0.007$) of existence of the quasicrystal phase
at p = 0.02. This value is in good agreement with the value ($\sqrt{3}/4$) 
obtained with random tiling models 
(which cover the plane with a set of rigid tiles without gap) when the fraction
of total tiling area occupied by square tiles is equal to the fraction 
occupied by triangle tiles, i.e when the system exhibit twelvefold rotational
symmetry. At the same reduced pressure, the ratio square 
tiles over triangle tiles is $\sim 0.41$ for the N = 100 nodes system.
Finite-size effects seems to decrease the probability of occurence
of geometrical defects.  

The model when both topological and geometrical defects are
allowed (Figure~\ref{fig:2d_diag}c) exhibits a richer phase behavior than
the only geometrical model with the presence of a hexagonal crystal (X$_h$) 
phase, a square crystal (X$_s$) phase, a dodecagonal quasicrystal (QX) phase 
and a liquid (L) phase. Typical examples of these phases are shown on
Figure~\ref{fig:2d_snap2}.  
Two triple points are present, a (QX, X$_s$, L) located around $p \simeq 0.03$
and $t \simeq 0.0065$ and a (X$_h$, QX, L) triple point located around
$p \simeq 0.06$ and $t \simeq 0.006$.
If we make abstraction of the presence of the square crystal phase and we
extend the upper phase boundary of the quasicrystalline phase toward $p = 0$,
we notice that the temperature range of existence of the quasicrystalline
phase is comparable to the one present in the only geometrical model.
However, the stability range in pressure of the quasicrystalline phase is 
almost twice as large as in the only geometrical model, due to the presence
of topological defects which allow a larger gain both in entropy of 
conformation (by introducing tiling faults) and in entropy of vibration.
By comparison with the only geometrical model, the number of square tiles
divided by the number of triangle tiles is around $\sim 0.45$ in the
quasicrytalline phase, at p = 0.02 for the system with N = 100 nodes.
At p = 0.01 and p = 0.04, for the N = 1024 system, the ratio square/triangle
is roughly in the range $\sim 0.43-0.44$.  
Quite surprisingly, for pressure lower than $p \simeq 0.03$, a square crystal
phase preempts, at high enough temperature, the full apparition of the 
quasicrystalline phase. This phase seems to become energetically favorable 
with respect to both the quasicrystalline phase and the liquid phase and 
quite interestingly is not present for the only geometrical model. 
It seems likely that this is due to a feature of the elastic model, in which
a bond entropy is contributing to the total entropy of the system. Because
a broken bond is free to flip without any energy cost, the gain in bond entropy 
overcomes the loss in entropy of configuration and the increase in volume and
stabilize the square crystal with respect to the quasicrystal and the liquid
phase at low enough pressure and high enough temperature. 
This is probably the same reason which leads, at the same pressure (p = 0.02)
and for the same temperature range, the quasicrystalline phase of the only 
goemetrical model to exhibit slightly less squarelike polygons than the 
topological and geometrical model. 
For pressures larger than $p \simeq 0.06$ a transition between an hexagonal
crystal and a dense liquid occurs. The change in symmetry of the system is
governed by the topological defects. Studies at $p = 0.08$ with the $N = 1024$
nodes system seem to indicate that the location of the solid-liquid phase 
boundary is not very sensitive to finite-size effects. 

The only topological phase diagram (Figure~\ref{fig:2d_diag}b) exhibits 
a hexagonal crystal phase (X$_h$) and a dense liquid due to the presence of
topological defects. 
A phase transition between these two phases occurs over the full range of
pressures studied. At each pressure, a well
defined peak in the specific heat appears at the transition but due to finite 
size effects it is difficult to probe the nature of the transition.
The phase diagram reported here has been computed 
upon heating. No significant hysteresis has been observed upon cooling. 
Studies carried out for the larger system $N = 1024$ at $p = 0.02$ and 
$p = 0.08$ indicate that the location of the phase boundary is shifted upward
due to finite-size effects. Typical examples of the liquid phase are presented
on Figure~\ref{fig:2d_snap2}, for a temperature close to the melting point
and for a temperature deeper into the liquid phase.  
It is interesting to notice, from the `slope' of the solid-liquid phase
transition line, using the Clapeyron relation, that our model lead to the
formation of a liquid denser than its solid phase. Even if this feature is
not uncommon for real liquids, it seems that in our case, this effect is 
related to the intrinsic nature of the discrete model.
Topological defects seem to create local compressible regions, increasing the
density in the system. On the phase diagram is also reported the predicted 
theoretical phase transition (see Appendix) from the Kosterlitz-Thouless (KT) 
model which corresponds to the unbinding of pairs of dislocations. It is 
thought to be the upper limit for the transition to occur if no other 
mechanism preempts it.
We notice that the KT prediction for our model is around two times larger than
the transition temperatures obtained from the elastic model simulation. 
However, our estimation have been done at $T=0$ and taking into account the 
topological defects will decrease sensitively the predicted values.

The most striking feature of the phase diagrams presented in 
Figure~\ref{fig:2d_diag} is the dramatic decrease (by a factor three to four)
in the solid-liquid transition temperature upon inclusion of geometrical 
(as well as topological) defects. Evidently, the geometrical defects 
significantly stabilize the liquid phase relative to the solid phase. 
This is the potential energy per node $u = U / N$ which mainly set the scale
of the solid-liquid transition temperature. At $p = 0.08$ and $N = 1024$
the difference in potential energy per node at the transition 
$\Delta u(t_c \simeq 0.023)$ for the only topological model is $\sim 0.012$
while for the topological and geometrical model, at $t_c \simeq 0.0065$ it
is $\sim 0.003$ i.e. a factor four smaller in magnitude and of the same order 
of magnitude than the observed decrease in transition temperature.  
Upon melting, the appearence and proliferation of defects in the only 
topological defects induces high strain on the elastic network. 
Geometrical defects present in the topological and geometrical model, by
allowing the release of this high strain in the elastic network, are the
main responsible (an increase in entropy coming from the bond entropy
contribution present in our model could also play a secondary role)
of the dramatic decrease of the melting temperature.
This observation is quite supportive of the idea that the geometrical defects
are the main responsible of the thermodynamical phase transition.
In our picture the topological defects could be only a side consequence of the
creation of the relevant excitations (geometrical) to the melting transition.
Moreover, the transition temperatures are similar, at a given pressure, to
those obtained in the case when only geometrical defects are allowed,
enhancing the idea that the melting is mainly driven by the geometrical
excitations.



Finally, we can estimate the relative variation in volume upon melting.
For the only topological model, $\Delta v / v_{S}$, with 
$\Delta v = v_{L} - v_{S}$, $v_{L}$ and $v_{S}$ being respectively the volume 
of the liquid and solid phase is of the order of $0.02$. It is significantly 
lower than the WCA simulations where $\Delta v / v_{S} \simeq 0.033$.
It is also possible to estimate the change of entropy on melting from the
condition for chemical equilibrium, $\Delta g = 0 = \Delta u - T \Delta s
+ P \Delta v$, where $g$ is the Gibbs free energy per particle, $u$ is the
internal energy per particle and $s$ is the entropy per particle. This leads
to

\begin{equation}
\Delta s = \frac{1}{T} (\Delta u + P \Delta v)
\label{entropy}
\end{equation}

We found $\Delta s$ which vary from $0.8 k_{B}$ at $p=0.1$ and decrease to
$0.55 k_{B}$ at $p=0$. The entropy variation is larger than the one
usualy found in WCA simulation, in DRP or in theory. The WCA simulation
give, for $T=0.6$, $\Delta s \simeq 0.4 k_{B}$ which is the same
order of magnitude than the results obtained in other 2D simulations, in DRP
and in theory.

For the topological and geometrical model, the relative variation of volume 
on melting is slightly more important than
for the case when only topological defects are allowed. We found
a value in better agreement with the WCA simulation, $\Delta v / v_{S}
\simeq 0.033$ (at $p = 0.08$ with $N = 1024$), due to the presence of the
square like polygons. Using equation ~\ref{entropy} we compute the variation
of entropy on melting and we found a value quite similar to the case with
only topological defects, i.e. $\Delta s \simeq 0.77$.
As mentioned before, the melting transition and crystal-expanded crystal
transition present in the only geometrical model occur at temperatures of the
same order of magnitude due to the presence of the geometrical fluctuation.
However, the slope of the (X$_h$-EX) boundary is steeper than the slope of
the (X$_h$-L) boundary. Using the relationship $dp/dt = \Delta s / \Delta t$
combined to the fact that the change in volume upon transition is similar
for both models, lead to a change of entropy on melting larger when 
topological defects are allowed. 

\section{Liquid Structures}

The structure of the dense liquid exhibited by the elastic model is
investigated in more detail with the larger system, $N = 1024$, at a reduced
pressure $p = 0.08$ for both the only topological model and the topological 
and geometrical model. This relatively high pressure have been chosen in 
order to avoid the complication in the comparison introduced by the presence 
of quasicrystalline an squarelike crystal structure. 
In order to study the microscopic structure of the dense liquid we chose
a reduced temperature $t_l$ such that $(t_l - t_c)/t_c \simeq Cst$, the constant
being the same for both the only topological model and the topological
and geometrical model.
In addition, we compare the dense liquid structure to the WCA liquid
structure obtained at a density ${\rho}_l$. 
Figure~\ref{fig:2d_corr} shows the radial pair correlation function and
the bond-orientational correlation function for the topological and 
geometrical model at a reduced temperature $t = 0.007$. The correlation
functions exhibits the typical behavior of a dense liquid with short-range
positional and bond-orientational order (within the finite-size effects). 

Upon the solid-liquid transition, the number of topological and geometrical
defects increase dramatically, as shown on Figures~\ref{fig:2d_order}a,b 
for the only topological model and the topological and geometrical model.
We can notice that the number of topological defects is significantly higher 
when bond breaking is allowed (Table~\ref{tab:coord_numb}). This is mainly
due to the fact that a broken bond is free to flip without energy penalty in
our model. This is a feature which is present in the particle-based system
for which a very slight motion of neighbor particles leads to the creation
of a defect quadrupole (see Figure~\ref{fig:2d_cartoon}). It could also be 
due to stress release or screening effects which 
favor the proliferation of topological defects at a lower energetical cost. 
 
For the topological and geomterical model, the number of broken bonds 
(and then the number of 
square-like geometrical defects) in the liquid phase is around 28\%. 
By comparison, the WCA liquid exhibits only 
15\% of broken bonds at ${\rho}_l = 0.80$ (the chosen density for our 
comparison), this number increasing deeper in the liquid phase 
(22\% at ${\rho}_l = 0.70$ \cite{matt_big}).

On Figures~\ref{fig:2d_snap}a and ~\ref{fig:2d_snap}b, we notice that the 
topological defects are 
highly correlated forming chains of alternating signs. The geometrical 
defects are also highly correlated and aggregate in different structures.
It is of interest to try to make a vertex classification in which vertices
are classified according to the sequence of polygons (squares and triangles)
present around a given vertex.
Square-triangle (ST) tiling are models used to describe liquids 
\cite{collins,kawa,yi} and more
recently some dodecagonal quasicrystals \cite{chen,leung,yang,kuo}.
There are four vertex type, hereafter referred as type A, B, C and D 
which are involved in ST plane tiling \cite{matt_big}.
Our results, displayed on Table~\ref{tab:type_vertex} show that vertex of 
type B and C are quite common in our model. In this sense, the local 
arrangment of 3 and 4-sides polygons appears to be strongly conditionned by 
tiling rules with a clear tendancy for the 3 and 4-sides polygons to form 
structure characteristics of ST tiling.
It is worth noticing that the vertex types listed in 
Table~\ref{tab:type_vertex} account respectively 
for 99.8\% and 66\% of the total vertex types present in
the elastic model (topological and geometrical) and the WCA liquid. The 
relatively low percentage obtained for the WCA liquid is
due to the fact that vertex types involving tiles with more than four sides 
are present (for example vertex types involving one hexagonal tile account
for 22\% of the total vertex types present).
Our model exhibits a tendancy to have more type B and C vertex
than type A and B. If the population of vertex D is also low in the case
of the WCA, the vertex A (corresponding to a perfect local solidlike region)
are more numerous than in the elastic network model. This trend is
is increased further by decreasing the overall network rigidity. 

Of course, there is also many violation of the tiling rule.
The vertex types E, F, G, H, and I correspond to tiling faults
and they account respectively for 25\%, 43\% and 27\% of the total vertex type
respectively for the elastic model and the WCA liquid. 
Vertex types E and F appear to be the most common tiling faults with 
type E most probable than type F for both the elastic model 
and the WCA liquid. 
From a vertex type point of view, the elastic model exhibits
features which are close from those of the WCA liquid.  
The elastic model comes closer to the WCA liquid but
the network rigidity seems still slightly too small to be able to capture one 
important feature of the WCA liquid, {\sl i.e} the fact that the most common
vertex remains the type A, since more type B and C are present in our model. 

We can see the dense liquid as a generalized tiling model in which the
tiles (triangles and squares) are deformable, leading to the creation of 
tiling faults.
A measure of the rate of deformation of a perfect ST tile can be obtain by 
computing the tiling charge associated to each vertex in our system.
This can be done by computing \cite{matt_big}:

\begin{equation}
c_{\alpha} = 6 \left ( \sum_{j = 1}^{n_{\alpha}} \left ( \frac{1}{2} 
- \frac{1}{p_{j}} \right ) - 1 \right )
\end{equation}

where the sum ranges over the $n_{\alpha}$ polygons which are around a vertex
$\alpha$, and $p_{j}$ is the number of side of the $jth$ polygon.

It is easy to verify that the `quantum' of tiling charge is $1/10$ in 
generalized tilings consisting of tiles having six of fewer sides, and
we are expressing the tiling charge in tenth. For example vertices
E and F have a tiling charge of $\pm 1/2$, corresponding to $\pm 5$ tenths.
Vertices corresponding to the perfect ST tilings, i.e, types A, B, C and D
have a zero tiling charge. 
In our model, due to the restriction we imposed on the nature of the
allowed geometrical defects, only tiling charge of strengths $0$, $\pm5$ and 
$\pm10$ are possible.
We notice that in our system (Table~\ref{tab:tiling_charge}), the larger 
tiling charges account only for $\sim$ 0.2\% of the total tiling charge,
suggesting that the local arrangments
of polygons are strongly conditioned by tiling rules. There is a strong
tendancy for polygons to aggregate into structures which minimize the tiling
charge around the vertices. 
Then, the geometrical defects have attractive, anisotropic interactions
(due to their shape) that cause them to aggregate in a way that the tiling
charge at a given vertex is minimized.
Then the emerging picture is that the solid-liquid transition is the
result of the prioliferation and condensation of the geometrical defects 
into grain-boundary like structure. 

Figures~\ref{fig:2d_cluster}a,b reveal extensive regions of sixfold order 
in the dense WCA liquid.
The typical size of these solidlike regions increases rapidly near the
freezing density. The spatial aggregation of topological defects, evident
on Figures~\ref{fig:2d_snap}a,b is directly related to the other prominent 
feature of the dense liquid shown on Figures~\ref{fig:2d_cluster}a,b i.e. 
the existence of the 
large solidlike regions which appear as rafts of nearly hexagonal Voronoi cells.
This dramatic spatial inhomogeneity is a consequence of the defect condensation
transition which induce the melting transition and which is seen as the 
qualitative key feature for a microscopic understanding of a 2D dense liquid.
The same behavior have been observed in the WCA liquid.

The solidlike clusters are interesting not only by themself but also and
mainly because they are a very important feature of the stucture of a dense 
liquid and then, will play an important role on the resulting properties of the
 liquid, in particular on the transport properties like viscosity.
Contemporary liquid state theory is incapable of predicting the detailed
characteristics of such solid-like fluctuations.

Figure~\ref{fig:2d_dist}a show an example, for the only topological model 
in the dense liquid phase, of the cluster size distribution
$n_{s} = N_{s}/N$ where $N_s$ is the number of cluster of size $s$ and N is 
the total number of particle in the system. Due to the usual normalization
used in percolation theory,
$\sum_s n_s = N_c/N$,$N_c$ being the total number of ordered clusters.
The solid clusters were identified by using the criterion that
the local six-fold order parameter satisfy $|{\psi}_6| \ge 0.75$.
We found that the whole distributions are quite well described by the
functional form:

\begin{equation}
n_{s} = A s^{- {\tau}_s} \exp{(-s/{\xi}_s)} \label{eq:size}
\end{equation}

This is the form roughly predicted by the Fisher droplet model \cite{fisher}
of condensation and which represents a special case of the scaling 
{\sl ansatz} used in percolation theory \cite{stauffer1,stauffer2}.
We have a power law behavior for small cluster size, and a crossover towards
an exponential behavior for large cluster size. The fit to this functional
form are quite good with however systematic deviation for large $s$.
This deviation at large $s$ is more significant in the topological and 
geometrical model than in the only topological model. 

For the topological only model, ${\tau}_s$ varies in the range 
$1.3-1.5$ decreasing with decreased temperature, while in the topological 
and geometrical model ${\tau}_s$ varies in the range $1.2-1.3$ and exhibits 
a relatively smaller variation with a decrease in temperature. 
The only topological model bears more features with the WCA fluid, which
exhibits values of ${\tau}_s$ in the range $1.2-1.5$ decreasing with
decreasing density. It is interesting to notice that single percolation 
models exhibit a constant value of ${\tau}_s$ for varying site or bond 
occupation probabilities.
Though the interpretation of ${\tau}_s$ is not clear, it is probably connected
to the geometry of the boundaries between ordered regions and then, to the 
tiling rules governing the dense liquid.

In the same way, for the topological only model, ${\xi}_s$ (which characterizes
the typical size of the ordered clusters, lies in the
range $20-150$, decreasing with decrasing temparature
while ${\xi}_s$ varies in the range
${\xi}_s = 5$ to  ${\xi}_s = 10$, decreasing with decreasing
temperature for the topological and geometrical model. 
In comparison, the WCA fluid exhibits a divergence near the
freezing transition. This divergence has been attributed to a finite-size
effects and is connected to the presence of large-size clusters which
will be of large but finite in an infinite system but which are in the
system used spanning the entire simulation box. Such effect doesn't seem
to happen in our case.

In conclusion, for a finite system, $n_s$ consists of two parts: 
a size-independant portion at small $s$ and a size-dependant part at large $s$
which develops near freezing.

More detailed informations can be obtain by computing the shape of 
the clusters by calculating their radius of gyration which is related 
to their size $s$ by :

\begin{eqnarray}
R_{g} & = &\sqrt{\frac{1}{n_s} \sum_{j=1}^{n_s} {({\bf r}_{j} - {\bf r}_{\mbox{cm}})}^2}  \nonumber \\ \nonumber \\
& \propto & s^{1/D_{f}} \label{eq:gyration}
\end{eqnarray}

where the sum runs over the $j$ particles at position ${\bf r}_{j}$ inside the
given cluster 
and ${\bf r}_{\mbox{cm}}$ is the center of mass of the cluster. 
$D_f$ is the fractal dimension of the cluster, a value of $D_f = 2$ or
close to 2 indicate a smooth rounded cluster while a significantly 
lower value indicates a rough cluster shape.
To perform this analysis and due to the periodic boundary conditions 
the spanning clusters are identified and discarded. 
Figure~\ref{fig:2d_dist}b shows the distribution function for the radius 
of gyration $R_g$ as a function of the cluster size $s$ for the only 
topological model and in the dense liquid phase. 
A fit to the functional form gives the range of variation of $D_{f}$ with
the temperature. In the only topological model $D_f$ varies in the range
$D_f = 1.5$ to $D_f = 1.75$ near the freezing point, while in the
topological and geometrical model, $D_f$ varies in a range from $D_f = 1.6$
to $D_f = 1.9$ near freezing.  
For comparison, the WCA liquid exhibit a fractal dimension $D_f$
varying in th range $D_f = 1.6$ to $D_f = 1.85$ near the freezing density.
In the three cases, the same general qualitative feature is present: the
solid clusters have a much smoother interface with the surrounding liquid 
near the freezing point than deeper inside the dense liquid phase.
In addition, on a more quantitative point of view, it seems that the
topological and geometrical model gives cluster shape in closer agreement
with the WCA liquid than the topological only case. In particular the
shape of the solid clusters is far smoother near freezing for the 
toplogical and geometrical model than for the only topological model.

\section{Conclusion}
We have study a simple elastic network model in order to be able to
probe the relative role of the two fundamental excitations that, 
in our opinion, are important for the melting transition. 
Phase diagrams obtained when only geometrical defects are 
present exhibits 2 crystal phases with hexagonal symmetry and different 
density as well as a dodecagonal quasicrystalline phase. When topological 
defects are allowed a transition between a solid and a liquid occurs. 
The topological and geometrical model exhibits in addition a rich phase 
behavior with hexagonal crystal, dodecagonal crystal, square crystal and
liquid phases. The main result is that the solid-liquid transition temperature
exhibited by the model in which geometrical defects are allowed (in addition
of topological defects) is decreased by a factor 3 to 4 with the transition
temperature exhibited in the model where geometrical defects are forbidden.
Evidently, geometrical defects stabilize the liquid phase with respect to
the solid phase by releasing the high strain present in the elastic network
when only topological defects are present.
This observation is quite supportive of the idea that if topological defects
are responsible for the loss of long range positional and bond-orientational
order, geometrical defects contribute significantly to the thermodynamics of 
the phase transition. 

\acknowledgments

This work was supported by NSF MRSEC Grant DMR 98-09555.

\appendix
\section*{Interaction parameters}

The KT temperature transition is expressed by:

\begin{equation}
k_{B}T = \frac{\overline{K}}{16 \pi}
\end{equation}

where $\overline{K}$ is the bare-elastic constant and can be expressed
as a function of the Lam\'e coefficients of the 2D solid \cite{lame}:

\begin{equation}
\overline{K} = \frac{4 \mu B}{\mu + B}
\end{equation}

where $B = -A (\partial P/\partial A)$ is the bulk modulus, and $\mu$ is the
shear modulus expressed by \cite{frenkel}:

\begin{equation}
\mu = -\frac{A'}{A} {(1 + \epsilon)}^{-1} {\cal P} {(1 + {\epsilon}^{T})}^{-1}
\end{equation}

where $A'$ is the deformed volume, $\epsilon$ is the strain tensor,
${\epsilon}^T$ its transposed and ${\cal P}$ is the microscopic
stress tensor defined as :

\begin{equation}
{\cal P}_{\alpha \beta} = \frac{1}{A'} \sum_{i} \sum_{j>i} {r_{ij}}_{\alpha}
{F_{ij}}_{\beta}
\end{equation}

with ${r_{ij}}_{\alpha} = {r_{i}}_{\alpha} - {r_{j}}_{\alpha}$ is the 
$\alpha$-component of the distance between particles $i$ and $j$  and
${F_{ij}}_{\beta}$ is the $\beta$-component of the force due to the particle 
$i$ on particle $j$.

In order to have a rough analytical idea of the magnitude of the predicted
KT transition we compute the Lam\'e coefficients in the framework of our
model at $T=0$, i.e. considering that we have a perfect 2D-triangular solid
lattice. Moreover we assume that the pure shear strain is infinitesimal.

At the first order in strain magnitude, the bulk and shear modulus are 
expressed in dimensionless variables by : 
 
\begin{eqnarray}
B & = & \frac{\sqrt 3}{2} \left( {1 + \frac{p}{\sqrt 3} }\right) \\ 
& & \nonumber \\
\mu & = &\frac{p}{2} + \frac{\sqrt 3}{2} \left( {1 + 3 K
\left( {1 + \frac{p}{\sqrt 3} }\right) 
}\right)
\end{eqnarray}

Then, the dimensionless KT transition temperature $t_{KT} = k_{B}T_{KT}/K_{r}
a^{2}$ is :

\begin{equation}
t_{KT} = \frac{p + \sqrt{3} \left(
{1 + 3 K {\left( {1 + \frac{p}{\sqrt{3}} } \right)}^{2}}
\right)}{1+ \frac{3}{2} K \left( {1 + \frac{p}{\sqrt{3}} } \right)}
\end{equation}


\newpage

\begin{figure}
\caption{
Voronoi construction for two nearly square arrangements of particles,
showing how a small displacement of the particles can create a disclination
quadrupole (right) from a configuration with no disclination. Schematic
representation of the holes created by a geometrical non topological defect
(left).
}
\label{fig:2d_cartoon}
\end{figure}

\begin{figure}
\caption{
Schematic representation of the discrete elastic model, in which
each node is connected to its nearest-neighbors with an elastic spring
of equilbriuum length $a$ and elastic constant $K_r$. In addition,
bond-angle springs with equilibrium bending angle ${\theta}_p$ and elastic
constant $K_{\theta}$ are used between ajacent springs.
}
\label{fig:2d_model}
\end{figure}

\begin{figure}[htb]
\caption{
Illustration of the MC moves performed on the springs (a) bond flipping to
mimic the creation of topological defects (arrising from the unbinding of
a quadrupole); (b) bond breaking to mimic the creation of a geometrical
defect. 
}
\label{fig:2d_moves}
\end{figure}

\begin{figure}
\caption{
Phase diagrams obtained for a system of $N = 100$ nodes with $K = 0.1$ as
a function of temperature $t$ and pressure $p$
(a) only geometrical defects; (b) only topological defects, the upper
dashed line being an estimate of the Kosterlitz-Thouless transition 
temperature; (c) both topological and geometrical defects.
$X_h$ represents an hexagonal crystal phase, $X_s$ a square crystal phase,
$QX$ a dodecagonal quasicrystal phase and $L$ the dense liquid phase.
}
\label{fig:2d_diag}
\end{figure}

\begin{figure}
\caption{
Representative evolution of various order parameters (top) and
defets distribution (bottom), as a function of the reduced temperature, for the
topological and geometrical model, with $N = 1024$ at p = 0.01.
Top:${|{\Psi}_4|}^2$ ($\bullet$), ${|{\Psi}_6|}^2$ ($\Box$); 
${|{\Psi}_{12}|}^2$ ($\triangle$);
Bottom: topological defects ($5$ and $7$-coordination nodes) 
($\bullet$); geometrical defects (broken bonds) ($\triangle$).
The vertical lines represent the location of X$_h$-QX, QX-X$_s$ and X$_s$-L
phase transitions. 
}
\label{fig:2d_order}
\end{figure}


\begin{figure}
\caption{
Representative configurations for the topological and geometrical model for
N = 1024.
From top to bottom and from left to right: hexagonal crystal (X$_h$) at
p = 0.08, t = 0.005; square crystal (X$_s$) at p = 0.01, t = 0.007; 
dodecagonal quasicrystal (QX) at p = 0.04, t = 0.0056; dense liquid (L)
at p = 0.08, t = 0.007. Red circle represent -1 topological defects
and blue circle represent +1 topological defects. Only non-broken bonds are 
represented.
}
\label{fig:2d_snap}
\end{figure}

\begin{figure}
\caption{
Representative configurations for the only geometrical model (top)
and for the only topological model (bottom), for N = 1024.
From top to bottom and from left to right: dodecagonal crystal (QX) at
p = 0.02, t = 0.0064; expanded crystal (EX) at p = 0.02, t = 0.008; 
dense liquid (L) at p = 0.08, t = 0.024 ($t_c \sim 0.023$); dense liquid (L)
at p = 0.08, t = 0.028. 
Red circle represent -1 topological defects
and blue circle represent +1 topological defects. Only non-broken bonds are 
represented. 
}
\label{fig:2d_snap2}
\end{figure}


\begin{figure}
\caption{
Typical radial correlation function in the dense liquid phase of the topological
and geometrical model, N = 1024, p = 0.08, t = 0.007.  
Top: radial pair correlation function g(r); Bottom: radial bond-orientational 
order correlation function g$-6$(r). 
}
\label{fig:2d_corr}
\end{figure}

\begin{figure}
\caption{
Distribution functions, in the case of the only topological model with 
$K = 0.1$ and $t = 0.025$, of (a) the cluster size $n_s$. The solid line 
represents a fit to Equation (18); (b) the cluster
radius of gyration. The solid line represents a fit to Equation (19).
}
\label{fig:2d_dist}
\end{figure}



\newpage
\newpage

\begin{table}[htb]
\begin{center}
\caption{Probalities of occurence of various coordination number}
\label{tab:coord_numb}
\begin{tabular}{cccc}
Coordination number  & topological and geometrical & only topological 
& WCA liquid \\  
\tableline
3     & 0 & 0 & 0 \\
4     & 0 & 0 & 4.31 $\times$ 10$^{-4}$\\
5     & 0.2284 & 0.1020 & 0.1513\\
6     & 0.5432 & 0.7960 & 0.6987\\
7     & 0.2284 & 0.1020 & 0.1469\\
8     & 0 & 0 & 2.63 $\times$ 10$^{-3}$\\
9     & 0 & 0 &2.8 $\times$ 10$^{-8}$ \\
\end{tabular}
\end{center}
\end{table}

\begin{table}[htb]
\begin{center}
\caption{Probalities of occurence of various vertex type}
\label{tab:type_vertex}
\begin{tabular}{ccc}
Vertex type  & Topological and geometrical & WCA liquid \\ 
\tableline
A     & 0.116  & 0.158\\ 
B     & 0.288  & 0.123\\
C     & 0.319  & 0.106\\
D     & 0.021  & 0.005\\
E     & 0.082  & 0.111\\
F     & 0.073  & 0.102\\
G     & 0.045  & 0.027\\
H     & 0.027  & 0.017\\
I     & 0.027  & 0.015\\
\end{tabular}
\end{center}
\end{table}

\begin{table}[htb]
\begin{center}
\caption{Probalities of occurence of various values of the tiling charge}
\label{tab:tiling_charge}
\begin{tabular}{ccc}
Tiling charge (in tenths)  & Topological and geometrical & WCA liquid \\ 
\tableline
-10     & 9 $\times$ 10$^{-3}$ & 5.67 $\times$ 10$^{-3}$\\ 
-9     & 0 & 2.25 $\times$ 10$^{-3}$\\
-7     & 0 & 0.0378 \\
-6     & 0 & 3.13 $\times$  10$^{-4}$\\
-5     & 0.1269 & 0.1462\\
-4     & 0 & 9.6 $\times$ 10$^{-3}$\\
-2     & 0 & 0.1056\\
0      & 0.7447 & 0.4017 \\
1      & 0 & 8.26 $\times$ 10$^{-3}$ \\
3      & 0 & 0.0901 \\
4      & 0 & 5.7 $\times$ 10$^{-4}$\\
5      & 0.1263 & 0.1413 \\
6      & 0 & 8.53 $\times$ 10$^{-3}$\\
8      & 0 & 0.0224 \\
9      & 0 & 1.16 $\times$ 10$^{-4}$\\
10     & 0.0012 & 0.0107
\end{tabular}
\end{center}
\end{table}

\newpage

\setcounter{figure}{0}

\newpage

\begin{figure}[htb]
\includegraphics[width=3.0in]{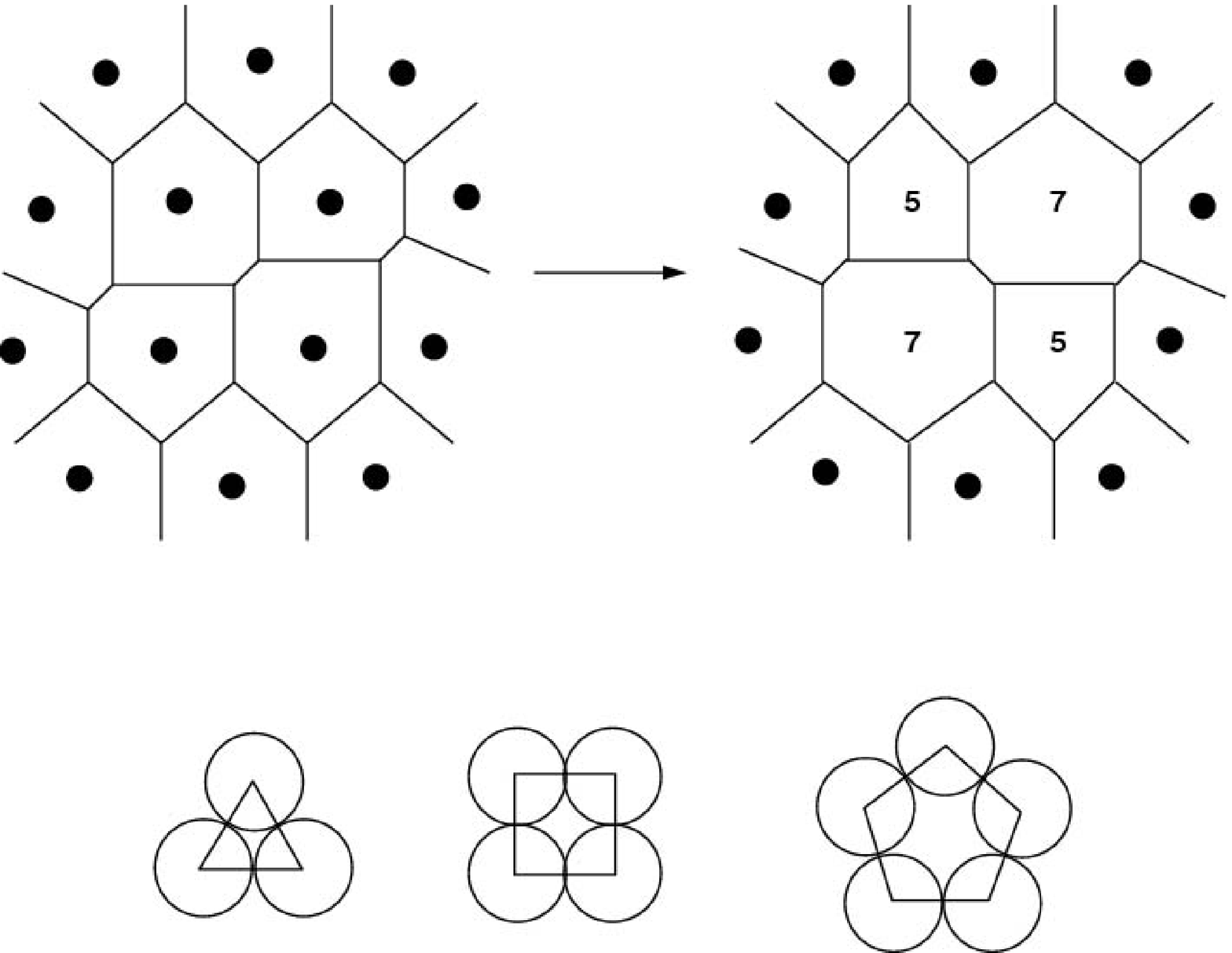}
\caption{ }
\end{figure}

\newpage

\begin{figure}[htb]
\includegraphics[width=3.0in]{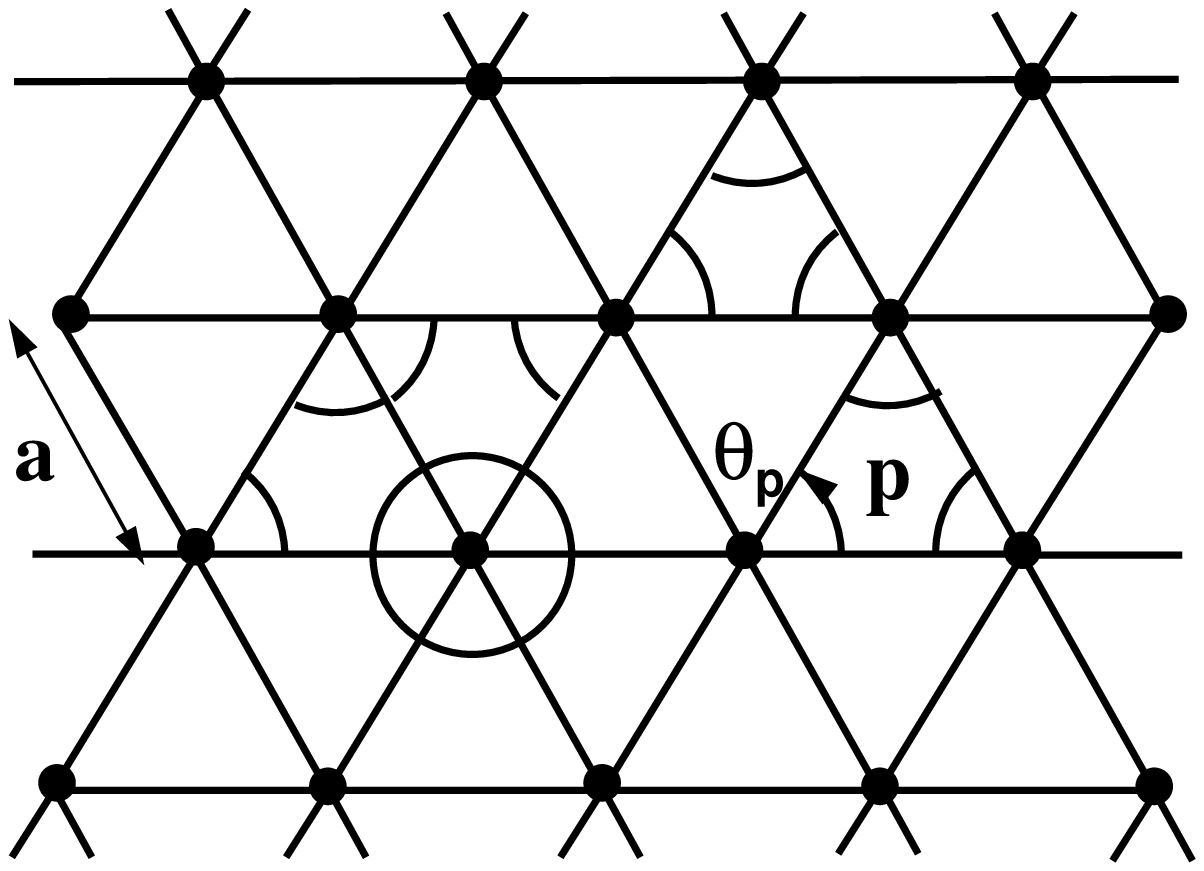}
\caption{ }
\end{figure}

\newpage

\begin{figure}[htb]
\includegraphics[width=3.0in]{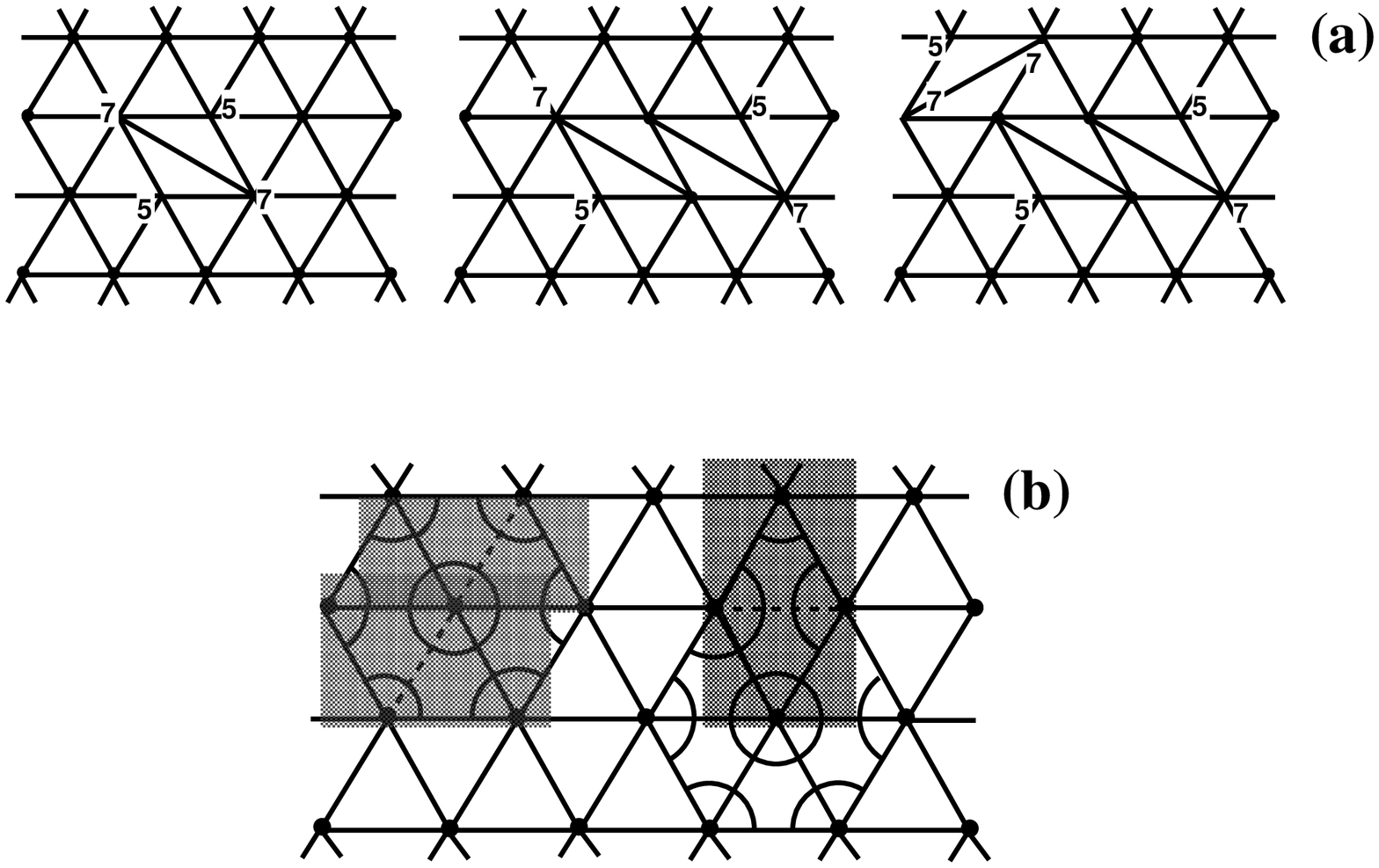}
\caption{ }
\end{figure}

\newpage

\begin{figure}[htb]
\includegraphics[width=3.0in]{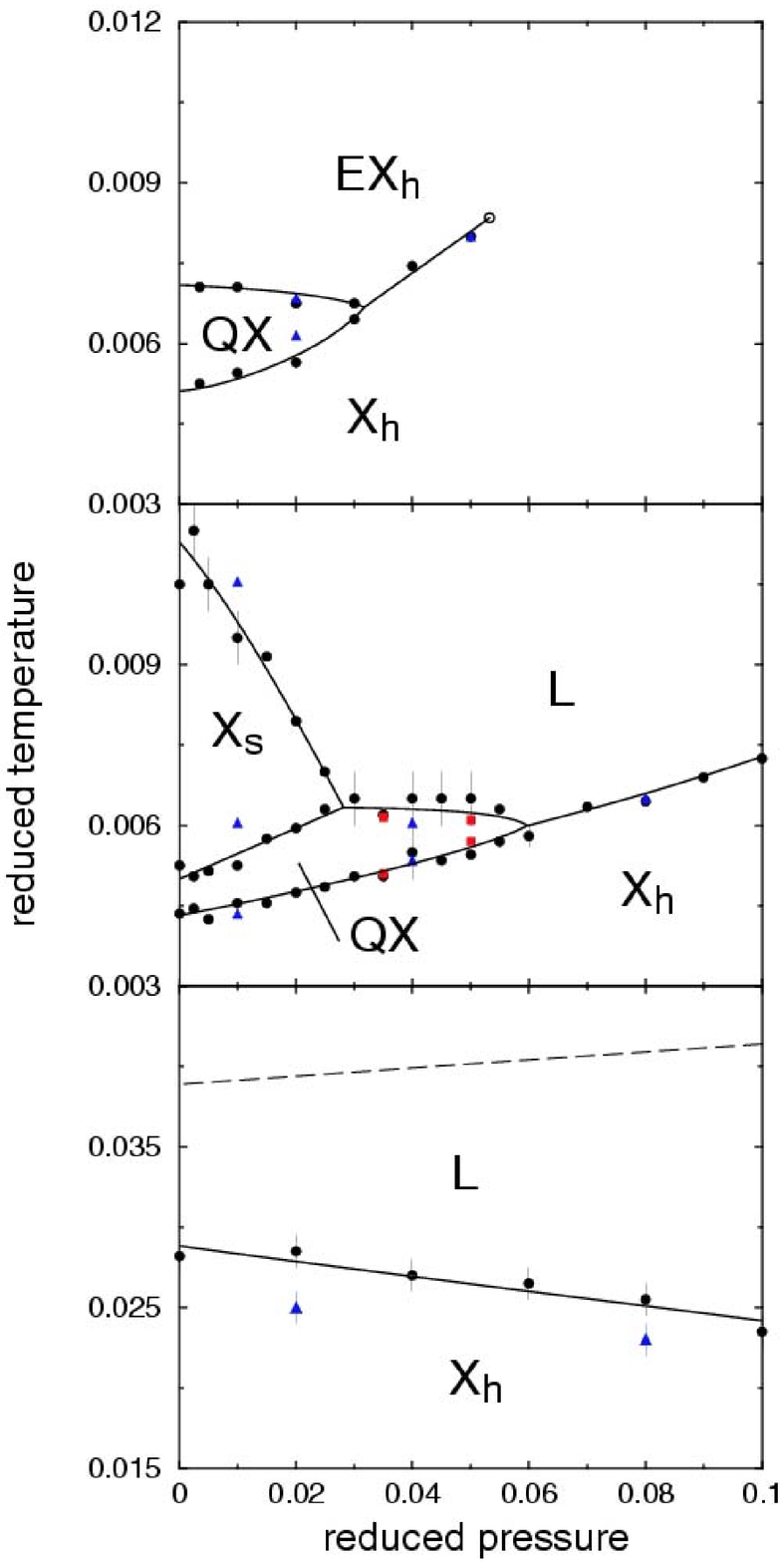}
\caption{ }
\end{figure}

\newpage

\begin{figure}[htb]
\includegraphics[width=6.0in]{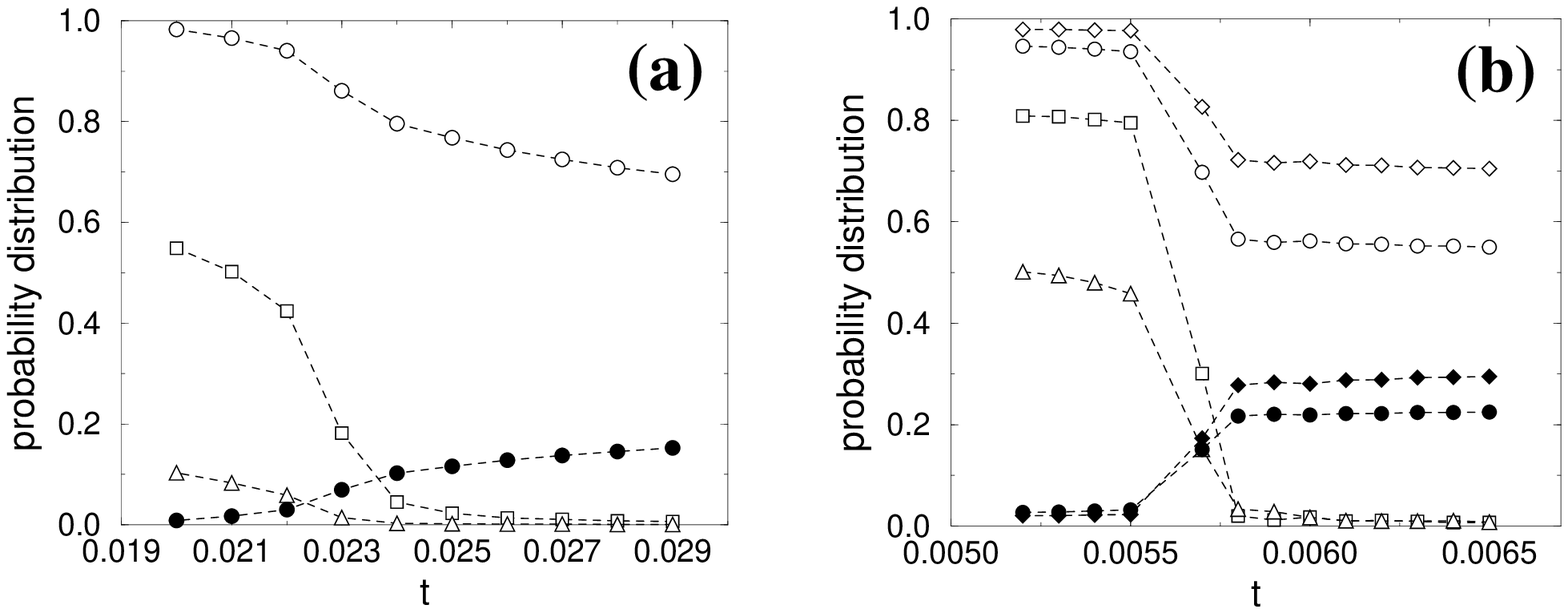}
\caption{ }
\end{figure}

\newpage

\begin{figure}[htb]
\includegraphics[width=5.0in]{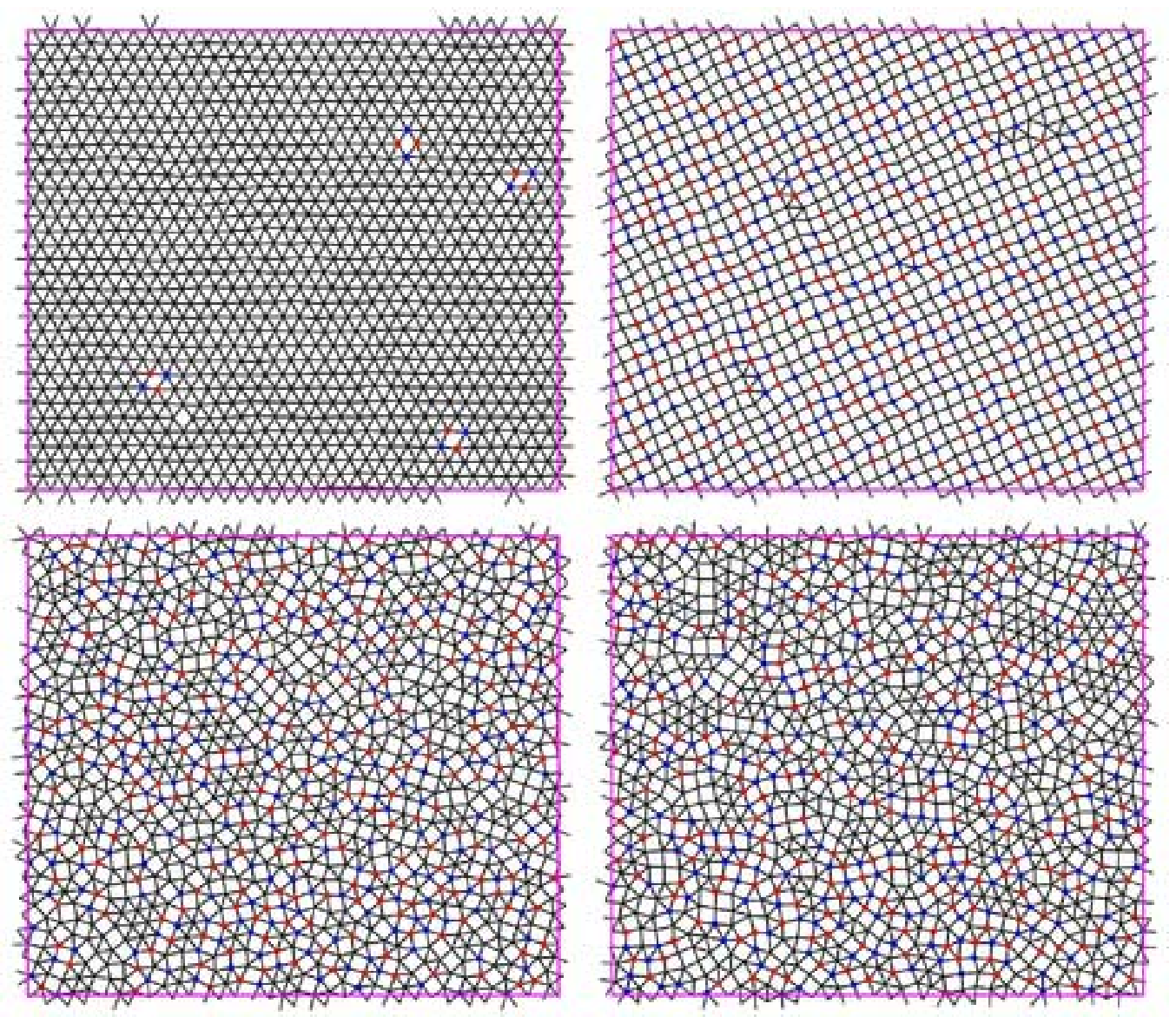}
\caption{ }
\end{figure}

\newpage

\begin{figure}[htb]
\includegraphics[width=5.0in]{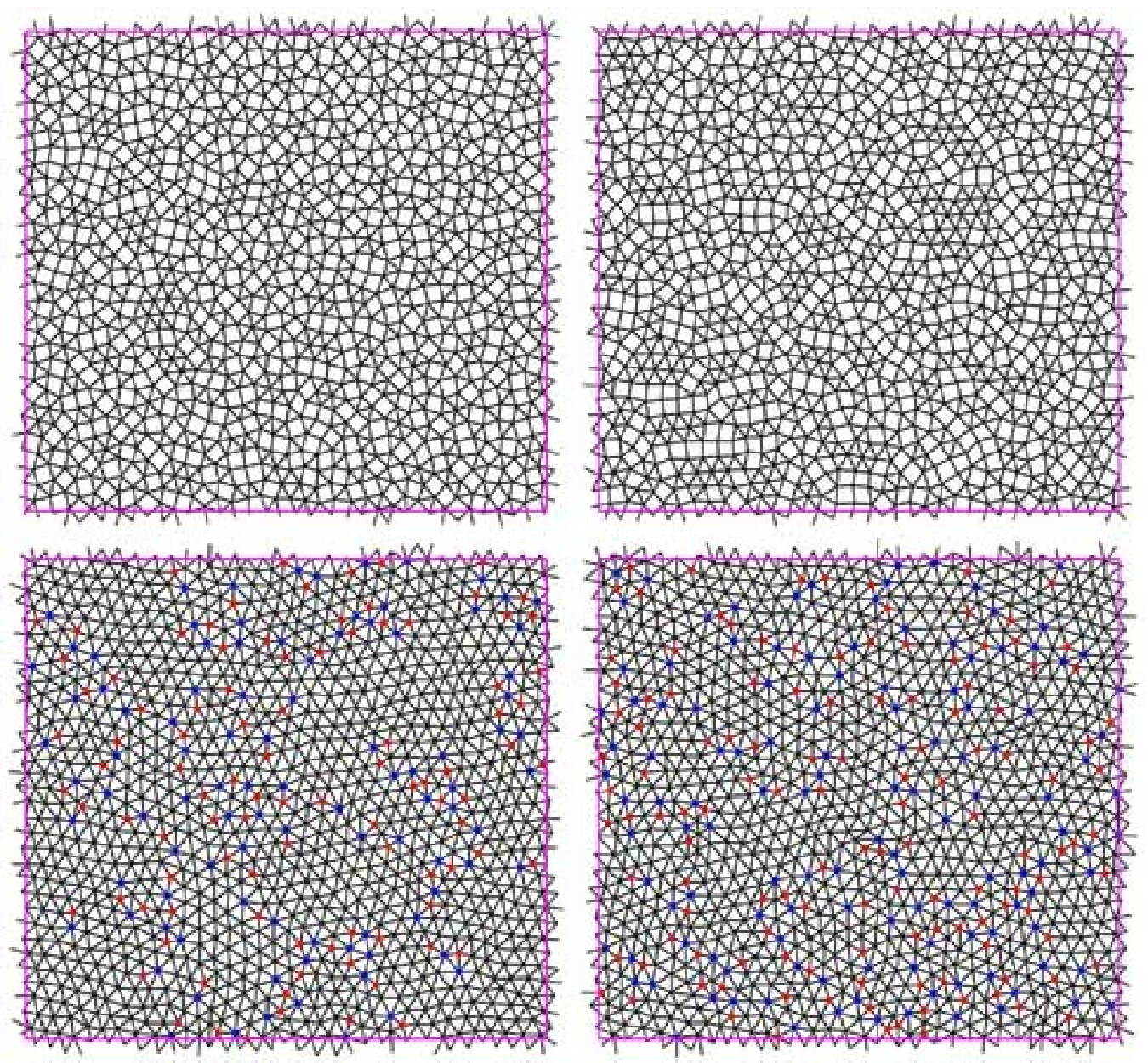}
\caption{ }
\end{figure}

\newpage

\begin{figure}[htb]
\includegraphics[width=3.0in]{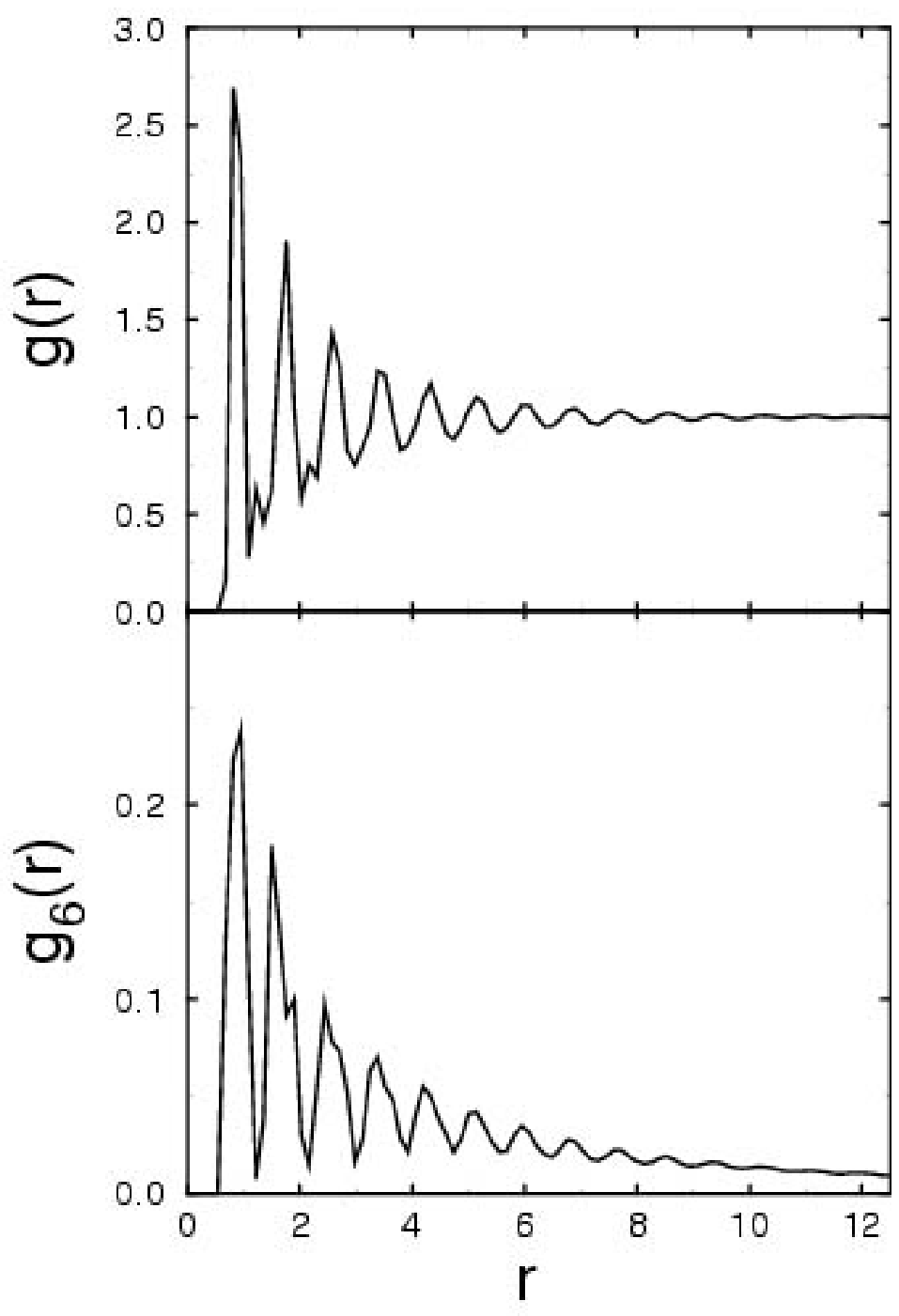}
\caption{ }
\end{figure}

\newpage

\begin{figure}[htb]
\includegraphics[width=5.0in]{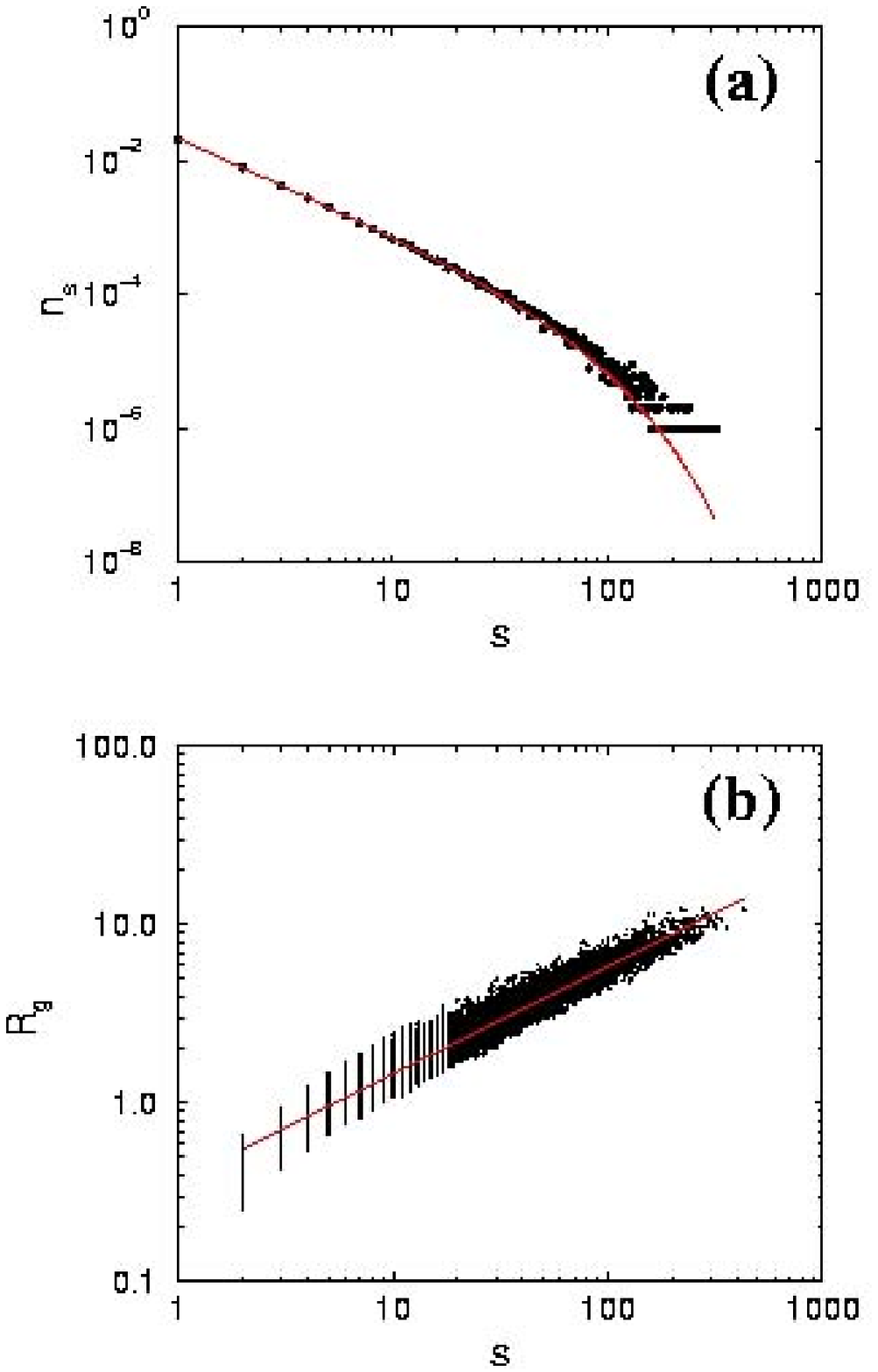}
\caption{ }
\end{figure}






\begin{references}

\bibitem{matt_big}
M.\ A. Glaser and N.\ A.\ Clark,
{\it Adv.\ Chem.\ Phys.}\ {\bf 83}, 543 (1993).

\bibitem{bern1}
J. \ D.\ Bernal,
{\it Nature}\ {\bf 183}, 141 (1959).

\bibitem{bern2}
J. \ D.\ Bernal,
{\it Nature}\ {\bf 185}, 68 (1960).

\bibitem{bern3}
J. \ D.\ Bernal,
{\it Proc.\ R. \ Soc.\ A}\ {\bf 280}, 299 (1964).

\bibitem{bern4}
J. \ D.\ Bernal,
in {\it Liquids: Structure, Properties, Solid Interactions}, 
T.\ J.\ Hughel, Ed.
(Elsevier, Amsterdam, 1965), p.25.

\bibitem{hansen}
J.\ P.\ Hansen and I. \ R.\ McDonald,
{\it Theory of Simple Liquids}, (Academic Press, London, 1986)

\bibitem{mermin1}
P. \ C. \ Hohenberg,
{\it Phys.\ Rev.}\ {\bf 158}, 383 (1967)

\bibitem{mermin2}
N. \ D. \ Mermin and H.\ Wagner,
{\it Phys.\ Rev. \ Lett.}\ {\bf 17}, 1133 (1966)

\bibitem{mermin3}
N. \ D. \ Mermin,
{\it Phys.\ Rev.}\ {\bf 176}, 250 (1968)

\bibitem{kost1}
J. \ M.\ Kosterlitz and D.\ J.\ Thouless,
{\it J.\ Phys.\ C}\ {\bf 5}, L124 (1972)

\bibitem{kost2}
J. \ M.\ Kosterlitz and D.\ J.\ Thouless,
{\it J.\ Phys.\ C}\ {\bf 6}, 1181 (1973)

\bibitem{halp1}
B.\ I.\ Halperin and D.\ R.\ Nelson,
{\it Phys.\ Rev.\ Lett.}\ {\bf 41}, 121 (1978)

\bibitem{halp2}
D.\ R.\ Nelson and B.\ I.\ Halperin,
{\it Phys.\ Rev.\ B}\ {\bf 19}, 2457 (1979)

\bibitem{young}
A.\ P.\ Young,
{\it Phys.\ Rev.\ B}\ {\bf 19}, 1855 (1979)

\bibitem{collins}
R.\ Collins,
{\it Proc.\ Phys.\ Soc.}\ {\bf 83}, 553 (1964)

\bibitem{kawa}
H.\ Kawamura,
{\it Prog.\ Theor.\ Phys.}\ {\bf 70}, 352 (1983)

\bibitem{yi}
Y.\ M.\ Yi and Z.\ C.\ Guo,
{\it J.\ Phys.:\ Condens.\ Matter}\ {\bf 1}, 1731 (1989)

\bibitem{ziman}
J.\ M.\ Ziman,
{\it Models of Disorder}\ (Cambridge University Press, Cambridge, England, 1979)

\bibitem{finney}
J.\ L.\ Finney,
{\it Mater.\ Sci.\ Eng.}\ {\bf 23}, 207 (1976)

\bibitem{voronoi}
G.\ F.\ Voronoi and J.\ Reine,
{\it Angew.\ Math.}\ {\bf 134}, 198 (1908)

\bibitem{berker1}
A.\ N.\ Berker and D.\ Anderman,
{\it J.\ Appl.\ Phys.}\ {\bf 53}, 7923 (1982)

\bibitem{berker2}
B.\ Nienhuis, A.\ N.\ Berker, E.\ K.\ Riedel and M.\ Schick,
{\it Phys.\ Rev.\ Lett.}\ {\bf 43}, 737 (1979)

\bibitem{metro}
N.\ Metropolis, A.\ W.\ Rosenbluth, M.\ N.\ Rosenbluth, A.\ N.\ Teller, and
E.\ Teller,
{\it J.\ Chem.\ Phys.}\ {\bf 21}, 1087 (1953)

\bibitem{chen}
H.\ Chen, D.\ X.\ Li and K.\ H.\ Kuo,
{\it Phys.\ Rev.\ Lett.}\ {\bf 60}, 1645 (1988)

\bibitem{leung}
P.\ W.\ Leung, C.\ L. Henley and G.\ V.\ Chester,
{\it Phys.\ Rev.\ B}\ {\bf 39}, 446 (1989)

\bibitem{yang}
Q.\ B.\ Yang and W.\ D.\ Wei,
{\it Phys.\ Rev.\ Lett.}\ {\bf 58}, 1020 (1987)

\bibitem{kuo}
K.\ H.\ Kuoi, Y.\ C.\ Feng and H.\ Chen,
{\it Phys.\ Rev.\ Lett.}\ {\bf 61}, 1740 (1988)

\bibitem{lame}
J.\ Friedel,
{\it Dislocations}\ (Pergamon Press, London, 1964)

\bibitem{frenkel}
D.\ Frenkel,
in {\it Simple Molecular Systems at Very High Density}\ edited
by A.\ Polian, P.\ Loubeyre, and N.\ Boccara (Plenum, New York, 1988)

\bibitem{fisher}
M.\ E.\ Fisher,
{\it Physics}\ {\bf 3}, 255 (1967)

\bibitem{stauffer1}
D.\ Stauffer,
{\it Phys.\ Rep.}\ {\bf 54}, 1 (1979)

\bibitem{stauffer2}
D.\ Stauffer,
{\it Introduction to Percolation Theory}\ (Taylor \& Francis, London, 1985)
\end{references}
\end{document}